\documentclass{article}
\usepackage{amssymb}
\usepackage{amsfonts}
\usepackage{amsmath}
 \usepackage{graphicx} 
\usepackage{hyperref}
\numberwithin{equation}{section}
\bibstyle{amsplain}

\newtheorem{theorem}{Theorem}[section]
\newtheorem{corollary}{Corollary}
\newtheorem{lemma}[theorem]{Lemma}
\newtheorem{proposition}{Proposition}

\newtheorem{definition}[theorem]{Definition}
\newtheorem{remark}{Remark}
\newcommand{\ep}{\varepsilon}
\newcommand{\eps}[1]{{#1}_{\varepsilon}}

\newcommand{\abs}[1]{\left\vert#1\right\vert}
\newcommand{\norm}[1]{\left\Vert#1\right\Vert}
\newcommand{\paren}[1]{\left(#1\right)}
\newcommand{\bracket}[1]{\left[#1\right]}
\newcommand{\inner}[2]{\left\langle #1, #2 \right\rangle}
\newcommand{\dx}{\partial_{x} }
\newcommand{\exist}{\mathrm{exist.}}
\newcommand{\cont}{\mathrm{cont.}}
\newcommand{\dist}{\mathrm{d}}

\newcommand{\h}[1]{{H^{#1}}}
\newcommand{\dt}{\partial_{t}}
\newcommand{\f}{\phi}
\newcommand{\calF}{\mathcal{F}}
\newcommand{\calN}{\mathcal{N}}
\newcommand{\calE}{\mathcal{E}}
\newcommand{\R}{\mathbb{R}}
\newcommand{\twodrowvec}[2]{\left( #1,\quad #2\right)}
\newcommand{\proof}{\noindent \textbf{Proof}:\quad}
\newcommand{\qed}{\begin{flushright}{$\blacksquare$}\end{flushright}}

\author{Gideon Simpson, Michael I. Weinstein, Philip Rosenau}
\title{On a Hamiltonian PDE arising in Magma Dynamics}

\begin{document}
\maketitle
\centerline{\scshape Gideon Simpson }
\medskip
{\footnotesize
 \centerline{Department of Applied Physics and Applied Mathematics, Columbia University}
   \centerline{New York, NY 10027, USA}
} 

\medskip

\centerline{\scshape Michael I. Weinstein }
\medskip
{\footnotesize
 \centerline{Department of Applied Physics and Applied Mathematics, Columbia University}
   \centerline{New York, NY 10027, USA}
} %
\medskip

\centerline{\scshape Philip Rosenau }
\medskip
{\footnotesize
 \centerline{School of Mathematical Sciences, Tel-Aviv University}
   \centerline{Tel-Aviv 69978, Israel}
} %

\bigskip

 \centerline{Address for correspondence: grs2103@columbia.edu}

\begin{abstract}
In this article we discuss a new Hamiltonian PDE arising from a class of  equations appearing in the study of magma, partially molten rock, in the Earth's interior.  Under physically justifiable simplifications, a scalar, nonlinear, degenerate, dispersive wave equation may be derived to describe the evolution of $\phi$, the fraction of molten rock by volume, in the Earth.  These equations have two power nonlinearities which specify the constitutive realitions for bulk viscosity and permeability in terms of $\phi$.  Previously, they have been shown to admit solitary wave solutions.  For a particular relation between exponents, we observe the equation to be Hamiltonian; it can be viewed as a generalization of the Benjamin-Bona-Mahoney equation.
  We prove that the solitary waves are nonlinearly stable, by showing that they are constrained local minimizers of an appropriate time-invariant Lyapunov functional. A consequence is an extension of the regime of global in time well-posedness for this class of equations to (large) data, which include a neighborhood of a solitary wave.  Finally, we observe that these equations have {\it compactons}, solitary traveling waves with compact spatial support at each time.\end{abstract}

\section{Introduction} 
\label{sec:intro}
Consistent, macroscopic models of magma, partially molten rock, in the Earth's interior can were developed in \cite{McKenzie84, Scott84, Scott86}, coupling the flow of the solid rock with that of the liquid via conservation of mass and momentum.  Under appropriate assumptions (small fluid fraction, no large scale shear motions, etc.)  the system may be reduced to a single scalar equation for the evolution of the fluid fraction, the porosity $\phi$, \cite{Barcilon86, Barcilon89, Wiggins95}, that admit solitary waves.  In one spatial dimension, the equation is 
\begin{equation}
\label{eq:magma}
\dt\f + \dx\paren{\f^n} - \dx\paren{\f^n\dx\paren{\f^{-m}\dt\f}}=0
\end{equation}
with the boundary conditions that $\f(x,t) \to 1$ as $x \to \pm \infty$.  The nonlinearity parameter $n$ is specified by a Darcy's Law relationship between the permeability, $K$, of the rock and its porosity of the form $K(\f) \propto \f^n$.  The other nonlinearity parameter, $m$, is related to the bulk viscosity, $\zeta$, of the porous rock, with $\zeta(\f) \propto \f^{-m}$.  It is expected that $ 2\leq n\leq 5$ and $0 \leq m \leq 1$.  A well-posedness theory for the initial value problem of  (\ref{eq:magma}) is developed in \cite{Simpson07}; see also section \ref{sec-preliminaries}.

In the article \cite{Nakayama92}, solitary traveling waves, $\f=\f_c(x-ct)$, of speed $c$ are shown to exist for \eqref{eq:magma} for any speed satisfying $c>n> 1$; $m$ may take any real value. In many problems, solitary waves are well-known to be important coherent structures, participating in key dynamic processes. In order to play this role, solitary waves must be dynamically stable. The most direct approach to the nonlinear dynamic stability of solitary waves is via a variational structure of the equations.
Unfortunately, \eqref{eq:magma} does not appear to have such a structure for the parameter ranges $(m,n)$, arising in the magma problem.  However, while not of present interest to the problem of magma we observe that when $n+m=0$, there is a Hamiltonian formulation:
\begin{eqnarray}
\label{eq:hamiltonian-form}
\dt\f &=& \bracket{I - \dx(\f^n\dx(\f^{-m}\cdot))}^{-1}\dx (-\f^n) = J \frac{\delta \mathcal{H}}{\delta\phi}\\
\label{eq:hamiltonian}
\mathcal{H}[\f] &=& \int \paren{\frac{1-\f^{n+1}}{n+1}+\f-1}dx\\
\label{eq:skewsym-op}
J=J_\f&=& [I - \dx(\f^n\dx(\f^{-m}\cdot))]^{-1}\dx,\ \ \ n+m=0
\end{eqnarray}

We will make use of the Hamiltonian structure in these cases to show that their solitary waves are \emph{orbitally stable}, {\it i.e.}  for data sufficiently close to a solitary wave, the corresponding solution, modulo a time-dependent spatial translation, will remain close to the solitary wave in $\h{1}(\R)$.  The general method of proof is well established and discussed in \cite{Benjamin72, Bona75, Weinstein85, Weinstein86, Bona94} for the Kortweg - de Vries (KdV), Benjamin - Bona - Mahoney (BBM),  and Nonlinear Schr\"odinger (NLS) equations, amongst others.

We note that in contrast to generalizations of the BBM, KdV, NLS equations to arbitrary power nonlinearity, solitary waves of \eqref{eq:hamiltonian-form} are nonlinear stable for {\it arbitrary} powers, $n$.  Currently this is established up to a numerical computation computation of the slope of the function $c\mapsto\calN[\phi_c]$; see Proposition \ref{prop:increasing-momentum}.
 
This stability result is also of significance for the global existence theory for  \eqref{eq:magma};  at present, no global existence in time result is known for the case $n+m=0$.  We are required to prove, in tandem with the nonlinear stability, that solutions for data in a neighborhood of a solitary wave exist for all time. Specifically, we note that \eqref{eq:magma} can potentially become a degenerate dispersive equations if $\f$ tends to zero. As made clear in the well-posedness results \cite{Simpson07} global existence in time is ensured by uniformly bounding the porosity, $\f$, away from zero. Solitary waves  are examples of solutions, whose porosity is uniformly bounded   away from zero. The strategy is to show, using spectral and variational arguments,  that initial data, in a  small $H^1$ neighborhood of the solitary wave, remain in a small neighborhood, therefore persists in being bounded away from zero, ensuring global existence and stability.

We proceed as follows.  In Section \ref{sec-preliminaries}, we review some of the basic mathematical properties of \eqref{eq:magma} on well-posedness theory and solitary waves, and state the our main results: Theorem \ref{thm:orbital} and Corollary \ref{cor:globalexist}. In Section \ref{sec-invariants-waves}, we review the constants of motion and their relation to the solitary waves. The proofs of the main results, Theorem \ref{thm:orbital} and Corollary \ref{cor:globalexist}, on orbital stability and global existence of data near a solitary wave solution, are given Section \ref{sec-stability-proof}.  In Section \ref{sec:compacton}, we note a relationship between our equations and those that have \emph{compacton} solutions, solitary waves with compact support, \cite{Rosenau06a}, and show that \eqref{eq:magma} also possesses such solutions.

Finally, we remark that  in a forthcoming paper, we prove the \emph{asymptotic} stability of solitary waves in the general case (arbitrary $m$ and $n$)  of \eqref{eq:magma},  of small amplitude, without any restriction on $m$ and $n$ . In fact, the Hamiltonian structure which we presently use in the case $n+m=0$ for\eqref{eq:magma} has  implications for the linear spectral theory and stability analysis in this work, via the Evans function (see, for example, \cite{Pego92}), an analytic function, whose zeros are points in the discrete spectrum of and {\it resonances} of the linearized spectral problem about the solitary wave.
 
\section{Background and Main Results}
\label{sec-preliminaries}
We first state the basic results on well-posedness of \eqref{eq:magma}; see \cite{Simpson07}, Theorems 2.12 and 2.13.
\begin{theorem}(Local Existence in Time \& Continuous Dependence Upon Data)

Let $\phi_{0}$ satisfy
\[\norm{\phi_{0}-1}_\h{k}\leq R\quad\mbox{and}\quad \norm{\frac{1}{\phi_{0}}}_{\infty}\leq\frac{1}{2\alpha} \]
for $R>0$, $1\ge \alpha >0$, and $k\ge 1$.
\begin{description}
  \item[\textbf{(a)}]There exists $T_\exist>0$, $T_\exist=T_\exist(R,\alpha)$, and $\phi-1\in C^{1}([0,T_\exist):H^{k}(\mathbb{R}))$, such that $\f$ is a solution to \eqref{eq:magma} with  $\phi(\cdot,t)\ge \alpha $ and $\norm{\phi(\cdot,t)-1}_{H^{k}}\leq2 R$ for $t<T_\exist$.
\item[\textbf{(b)}]There is a maximal time of existence, $T_{\max}>0$, such that if
$T_{\max}<\infty$, then
        \begin{equation}
                \label{eq:hkblowup}
                        \lim_{t\to T_{\max}}\norm{\phi(\cdot,t)-1}_{H^{k}}+\norm{\frac{1}{\phi(\cdot,t)}}_{\infty}=\infty
        \end{equation}
\item[\textbf{(c)}]
Let $\phi^{(1)}$, $\phi^{(2)}$ be two solutions of \eqref{eq:magma}, existing in a common space $C^{1}([0,T):H^{k}(\mathbb{R}))$, $T>0$, and satisfying the bounds
\[
\norm{\f^{(j)}(t) -1}_\h{k}\leq 2 R \quad\mbox{and}\quad\norm{\frac{1}{\phi^{(j)}(t)}}_{\infty}\leq\frac{1}{\alpha}
\]
for some $0< \alpha\leq 1$, $R>0$, $k\geq1$, $j=1,2$, and all $t< T$.  Then there exists a constant, $K_\cont=K_\cont(R,\alpha,k)$, such that for any $t$ and $t'$,  $0\leq t \leq t' < T$,
\begin{equation}
\norm{\f^{(1)}(t') - \f^{(2)}(t')}_\h{k} \leq e^{K_\cont (t'-t)} \norm{\f^{(1)}(t)-\f^{(2)}(t)}_\h{k}
\label{eq:cont-dep}
\end{equation}
\end{description}
\label{thm:local-existence}
\end{theorem}

We will show that the solitary waves of \eqref{eq:magma} are \emph{orbitally stable} in the following sense.  Let us define the distance function,
\begin{definition}
Let $f-1$ and $g-1$ be in $\h{1}(\R)$.  Define the \emph{sliding metric} on $\h{1}$, $\dist$, 
\begin{equation}
\dist(f,g) = \inf_{y} \norm{f(\cdot+y)-g}_\h{1}
\label{eq:distance}
\end{equation}
\end{definition}
\begin{definition}
Given $\f-1\in C^{1}([0,T):H^{1}(\mathbb{R}))$,  $T>0$, we say that $\f$ is \emph{orbitally stable}, if for all $\eps>0$, there exists $\delta>0$ such that if
\[
\norm{\f(t=0)- \psi(t=0)}_\h{1}\leq \delta
\]
with $\psi-1 \in C^{1}([0,T):H^{1}(\mathbb{R}))$
\[
\dist(\f(t),\psi(t))<\ep
\]
for $t< T$.
\end{definition}

\begin{figure}
\begin{center}
\includegraphics[width=4in]{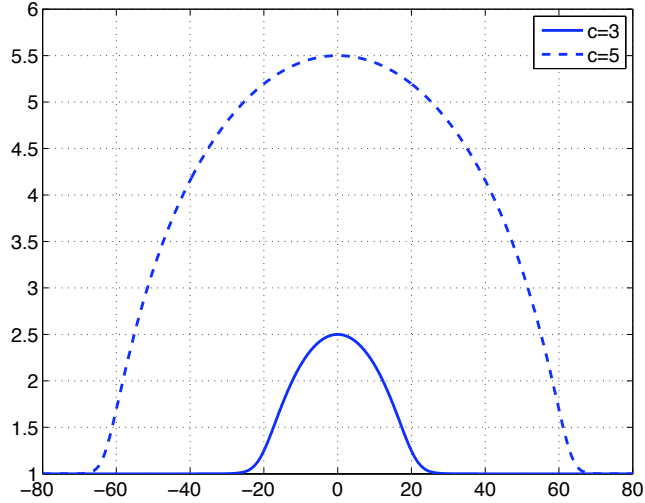}
\caption{Two solitary waves for $n=2$.  Note that as the speed parameter, $c$, increases, the waves become both taller \emph{and} broader.}
\label{fig:solwaves}
\end{center}
\end{figure}

A {\it solitary traveling wave} is a solution of the form $\f=\f_c(x-ct)$, where $\f_c(x)$ asymptotes to a constant, say $\f\equiv1$, as $x\to\pm\infty$.
Thus, solitary waves, $\f_c$, of \eqref{eq:magma}, for the case of $n+m=0$ satisfy
\begin{equation}
\label{eq:solitarywave1}
-c\dx \f_c +\dx(\f_c^n) + c \dx\paren{\f_c^n \dx(\f_c^n \dx\f_c) }=0
\end{equation}
After one integration, and using the boundary condition at $\infty$,
\begin{equation}
\label{eq:solitarywave2}
-c\f_c +\f_c^n + c \f_c^n \dx(\f_c^n\dx\f_c) +c-1=0
\end{equation}
In general, there is no closed form expression for $\f_c$ as a function of $x$.  As previously noted, for $c>n>1$, a solution in excess of the reference state at $\infty$ exists, \cite{Nakayama92}.  $\f_c(x)$ can be shown to be exponentially decaying as $x\to\pm\infty$ and analytic in a strip about the real axis in the complex plane, see \cite{Simpson07b} for more details.  Two such waves are pictured in Figure \ref{fig:solwaves}.

It is worth noting, that in \cite{Nakayama91, Nakayama94}, solutions for $c=n$ were found.  However, $\f_c(x)=0$ at isolated points and do not fit into our existence theory which relies on boundedness away from zero; hence, we do not consider them here.

\eqref{eq:solitarywave1} also possesses new compacton solutions, discussed in Section \ref{sec:compacton}.

Our Main Theorem and Corollary \ref{cor:globalexist} apply to the Hamiltonian case of \eqref{eq:magma}, $n+m=0$.
\begin{theorem}(Orbital Stability)

Let $\f_c$ be a solitary wave with $c>n$ and let $\f-1 \in C^1([0,T); \h{1}(\R))$ be a solution to \eqref{eq:magma}, $T>0$.  There exists  $\ep_\star=\ep_\star(\f_c)$ such that for all $\ep\leq \ep_\star$, there is a $\delta$ such that  if for some $x_0\in \R$,
\[
\norm{\f(\cdot, t=0)-\f_c(\cdot-x_0)}_\h{1} < \delta
\]
then $\dist(\f(t),\f_c) <\ep$ for all $t \in [0,T)$.
\label{thm:orbital}
\end{theorem}
We make no assumptions about $T$ in this theorem except that $T>0$; indeed it may be infinite.

\begin{corollary}(Global Existence and Orbital Stability)
Given a solitary wave $\f_c$, and $\ep\leq\ep_\star$, there exists $\delta>0$ such that if 
\[
\norm{\f_0-\f_c(\cdot-x_0)}_\h{1}\leq \delta
\]
with $x_0\in \R$, then $\f-1 \in  C^{1}([0,\infty):H^{1}(\mathbb{R}))$ and $\dist(\f(t),\f_c) < \ep$
for all time.
\label{cor:globalexist}
\end{corollary}

\section{Conservation Laws and Variational Characterization of Solitary Waves}
\label{sec-invariants-waves}
\subsection{Invariants and Regularity}
In addition to the Hamiltonian $\mathcal{H}$, another invariant is the generalized momentum
\begin{equation}
\label{eq:momentum}
\mathcal{N}[\f]=\int \paren{\frac{1}{2} \f^{2n}\f_x^2 +\frac{1}{2}(\f-1)^2}dx
\end{equation}
This was identified in \cite{Simpson07} as a well defined quantity for $\f-1\in\h{1}$, formed out of a linear combination of conservation laws discovered in \cite{Harris96}.  In Appendix \ref{sec:momentum}, we show the relationship between \eqref{eq:momentum} and the Lagrangian of \eqref{eq:magma}.

In order to prove that $\mathcal{N}$ and $\mathcal{H}$ are constant in time for $\h{1}$ solutions, one must first establish their conservation in $\h{2}$, which is obvious, and then approximate an $\h{1}$ solution in $\h{2}$ to show that these quantities really are invariant.  The time $T_\exist>0$ appearing in Theorem \ref{thm:local-existence} is chosen such that both $\mathcal{H}$ and $\mathcal{N}$ are invariant for $\h{1}$ solutions to \eqref{eq:magma}.  See Sections 4 and 5 of \cite{Simpson07} for details on the invariance of $\mathcal{N}$, which can easily be extended to $\mathcal{H}$.

For $\f-1 \in \h{1}$, $\mathcal{N}$ is obviously well defined; $\mathcal{H}$ is also well defined; the polynomial in the integrand $p(x) = (1-x^{n+1})/(n+1)+x-1$, has $p(1)=p'(1)=0$ and $p''(1)=-n$, giving the bound 
\[
\abs{\mathcal{H[\f]}}\leq C\norm{\f-1}_{L^\infty} \norm{\f-1}_\h{1}^2\leq C \norm{\f-1}_\h{1}^3
\]
with $C$ independent of $\f$.

\subsection{Variational Characterization of the Solitary Waves}
\label{sec:variational}
Let 
\begin{eqnarray}
\mathcal{E}[\f] &=& \mathcal{H}[\f] + c \mathcal{N}[\f]\nonumber\\
&=&\int \paren{\frac{1-\f^{n+1}}{n+1} + \f-1}dx + c \int\paren{\frac{1}{2}\f^{2n}\paren{\dx\f}^2 + \frac{1}{2}\paren{\f-1}^2}dx
\end{eqnarray}
For $c>n$, consider the taylor expansion of $\calE$ about a solitary wave $\f_c$,
\begin{equation}
\label{eq:expansion}
\mathcal{E}[\f_c + u] = \mathcal{E}[\f_c] + \left\langle \frac{\delta \mathcal{E}}{\delta\phi}[\f_c], u \right\rangle+ \frac{1}{2}\left\langle \frac{\delta^2 \mathcal{E}}{\delta\phi^2}[\f_c]u, u \right\rangle + O(\norm{u}_\h{1}^3)
\end{equation}
The variational derivatives are
\begin{eqnarray}
\label{eq:firstvariation}
\frac{\delta \mathcal{E}}{\delta\phi}[\f] &=& c \f -\f^n - c \f^n\dx\paren{\f^n \dx \f} -c+1\\
\label{eq:secondvariation}
L_c u &\equiv \frac{\delta^2 \mathcal{E}}{\delta\phi^2}[\f]u&=-c \dx \paren{{\f}^{2n} \dx u} - \bracket{(2n-1) n c\f^{2n-2}(\dx\f)^2+2 n c{\f}^{2n-1}\dx^2\f +n {\f}^{n-1} -c}u\nonumber \\ \label{eq:lin-op1}
&=& -c\f^n\dx^2\paren{\f^n u}+ \bracket{c -n \f^{n-1} - n c \f^{n-1}\dx\paren{\f^n \dx \f}} u \label{eq:lin-op2}
\end{eqnarray}
$\frac{\delta \mathcal{E}}{\delta\phi}[\f_c]=0$ because of \eqref{eq:solitarywave2}; a solitary wave of speed $c$ is a critical point of the this functional.  Alternatively, it can be viewed as critical points of $\mathcal{H}$, subject to the constraint of $\mathcal{N}$ with Lagrange multiplier $c$.

Since the solitary waves are critical points of $\calE$, we would \emph{like} to be able to make an analysis of the form
\[
\abs{\Delta \mathcal{E}} \geq\frac{1}{2}\inner{ L_c u} {u} \geq C\norm{u}_\h{1}^2 + O\paren{\norm{u}_\h{1}^3},\quad C>0
\]
to conclude their Lyapunov stability.  However, as proved in Proposition \ref{prop:Lc} , the Sturm-Liouville like operator, $L_c$, is not positive definite; it possesses a negative and a zero eigenvalue.  Nevertheless, there are natural constraints associated with the problem that will ensure positivity; these will be discussed in Section \ref{sec:constraints}.

For later use, we state a formal result on \eqref{eq:expansion}
\begin{lemma}
Given a Solitary Wave $\f_c$, there exist constants $D_2$ and $D_3$ such that for all such $u\in \h{1}$ with  $\norm{u}_\h{1}\leq \frac{1}{2}$.
\begin{eqnarray}
\label{eq:energybound1}
\abs{ \calE[\f_c+u]-\calE[\f_c] -\frac{1}{2}\inner{L_c u}{u}}&\leq& D_3\norm{u}_\h{1}^{3}\\
\label{eq:upper-polynomial}
\abs{\calE[\f_c+u]-\calE[\f_c]}& \leq &D_2 \norm{u}_\h{1}^2 + D_3 \norm{u}_\h{1}^3\equiv p_+\paren{\norm{u}_\h{1}}
\end{eqnarray}
\label{lem:energybound}
\end{lemma}
The polynomial $p_+$ will be used in the proof of the main theorem in Section \ref{section:orbital-proof}.

Another property of the invariant $\calN$ evaluated at $\f_c$ will imply that solitary waves of \eqref{eq:magma} are never unstable, see Theorem \ref{thm:orbital}.  
\begin{proposition}(Analytically confirmed for $n=2$, suggested numerically  $n\neq2$)
\label{prop:increasing-momentum}
\begin{equation}
\boxed{\frac{d}{dc}\mathcal{N}[\f_c] >0}
\label{eq-increasing-momentum}
\end{equation}
for all $c>n>1$.
\end{proposition}
\proof
Integrating \eqref{eq:solitarywave2} again, and applying the boundary condition at $\infty$, 
\begin{equation}
\label{eq:solitarywave3}
\frac{1}{2}\f_c^{2n} \paren{\dx\f_c}^2 = \frac{1}{2}\paren{\f_c-1}^2 +\frac{1}{c}\bracket{\f_c -1 +\frac{1-\f_c^{n+1}}{n+1}} = F(\f_c;c,n)
\end{equation}
Using this, and the even symmetry of $\f_c$, 
\[
\calN[\f_c] = 2 \int_0^\infty F(\f_c; c,n) + \frac{1}{2}\paren{\f_c-1}^2 dx
\]
\eqref{eq:solitarywave3} can be used to compute $\f_c^{\max}$ by solving $F\f_c^{\max};c,n)=0$.  Furthermore, we can compute $dx/d\f_c$ and make a change of variables with it to get
\begin{equation}
\label{eq:momentum-solitarywave}
\calN[\f_c] = 2 \int_1^{\f_c^{\max}} \bracket{F(y;c,n) + \frac{1}{2}\paren{y-1}^2}y^n \bracket{2 F(y;c,n)}^{-1/2}dy
\end{equation}

\begin{figure}
\begin{center}
\includegraphics[width=3.5in]{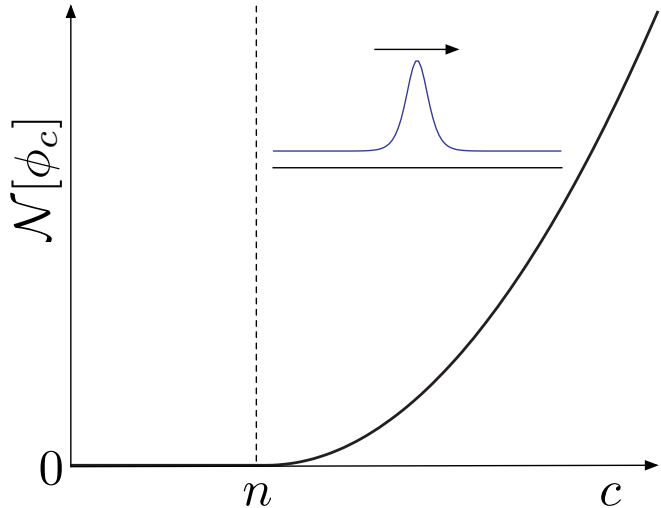}
\caption{A diagram of $\calN[\f_c]$.  $\calN[\f_c]$ and $d/dc \calN[\f_c]$ are zero at $c=n$. For $c\leq n$, there are not solitary waves.}
\label{fig-momentum-cartoon}
\end{center}
\end{figure}

At present, we have only been able to evaluate \eqref{eq:momentum-solitarywave} analytically in the case $n=2$.  Using \emph{Mathematica}, we compute
\begin{eqnarray*}
\f_c^{\max}&=&\frac{3c -4 }{2}\\
\mathcal{N}[\f_c] &=& \frac{12 \gamma ^3 \left(7 \gamma ^6+93 \gamma ^4+105 \gamma ^2+35\right)}{35 \left(\gamma ^2-1\right)^4} \quad\gamma = \sqrt{\frac{c-2}{c}}
\end{eqnarray*}
By inspection, this is strictly increasing for $c>2$.  The general behavior, both in this case and the rest, is diagrammed in Figure \ref{fig-momentum-cartoon}.  

\begin{figure}
\begin{center}
\includegraphics[width=5in]{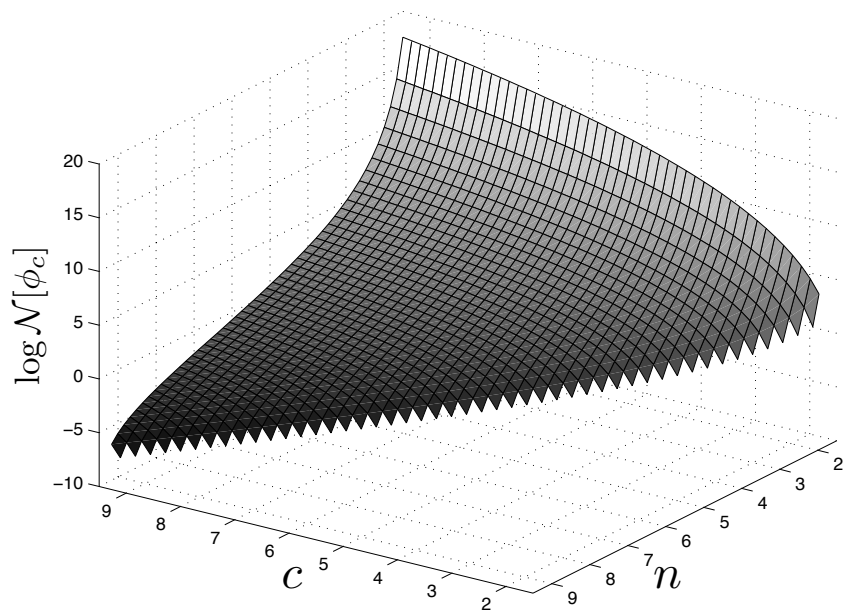}
\caption{A plot of $\log\mathcal{N}[\f_c]$, as a function of both $c$ and $n$, $c>n>1$, for the set $(c,n) \in [1.5,9.5]\times[1.5,9.5]$.  It is monotonic in both arguments, and in particular, increasing in $c$ for fixed $n$.}
\label{fig-functional-manifold}
\end{center}
\end{figure}

For $n\neq 2$, we justify our result with a computer plot, shown in Figure \ref{fig-functional-manifold}.  The manifold, $\mathcal{M}=(n,c,\mathcal{N}[\f_c])$, is monotonically increasing in $c$ for fixed $n$.  This was computed by first solving $F(\f_c^{\max};c,n)=0$ for $\f_c^{\max}$ using Brent's Method.  \eqref{eq:momentum-solitarywave} was then integrated using the QUADPACK routine QAWS to handle the $(\phi_c^{\max}-y)^{-1/2}$ singularity.  We used the GNU Scientific Library (GSL) implementation of these two methods, \cite{Galassi06}.
\qed

\section{Orbital Stability and Global Existence}
\label{sec-stability-proof}
\subsection{Constraints of the Flow}
\label{sec:constraints}
Under appropriate restrictions on our function space, $L_c$ will be a positive definite operator, allowing us to conclude orbital stability.  This is accomplished via two constraints discussed in the following two sections, \ref{sec:momentum-invariant} and \ref{sec-sliding-metric}.

\subsubsection{The $\mathcal{N}$ Invariant}
\label{sec:momentum-invariant}
We would like to assume $\mathcal{N}[\f] = \mathcal{N}[\f_c]$, and use this as a constraint, as Benjamin did in \cite{Benjamin72}.  This is of course not true for arbitrary $\h{1}$ perturbations.  Following, \cite{Bona75}, we show that for $\f$ sufficiently close to $\f_c$, we can find nearby solitary wave $\f_{c'}$ such that $\mathcal{N}[\f] = \mathcal{N}[\f_{c'}] $.

\begin{lemma}
\label{lem:nearby-wave}
Given a solitary wave $\f_c$, there exists $\delta_{c}>0$ such that for $\norm{u}_\h{1}\leq \delta_{\calN}$ there exists a $c'$ such that $\mathcal{N}[u+\f_c] = \mathcal{N}[\f_{c'}]$.  Furthermore, there exists a constant, $K_c = K_c(\f_c)$ such that $\abs{c-c'}\leq K_c \norm{u}_\h{1}$.
\end{lemma}
\proof This is proved using the implicit function theorem.  Let $\calF$ be the functional $\calF:  \h{1}(\R) \times \R\to \R$
\begin{equation}
\mathcal{F}[u,\Delta c] = \calN[\f_c +u]-\calN[\f_{c+ \Delta c}]
\end{equation}
$\calF[\mathbf{0},0]=0$ and the Fr\'echet derivatives are
\begin{equation}
D\calF[\bold{0},0]=\twodrowvec{\inner{c^{-1}(\f_c^n-1)}{\cdot}}{-\frac{d}{dc}\mathcal{N}[\f_c]}
\end{equation}
Both are bounded operators, and by Proposition \ref{prop:increasing-momentum} ,$\partial_c\mathcal{N}[\f_c]\neq 0$.  We may therefore apply the implicit function theorem for Banach spaces (see \cite{Reed80} for example), to conclude existence of an $\delta_{\calN}>0$ and a $C^1$ mapping $\mathcal{G}: \h{1} \to \R$, such that if $\norm{u}_\h{1}< \delta_{\calN} $, then $\mathcal{F}[u, \mathcal{G}[u]]=0$.  We then set $c' = c +  \mathcal{G}[u]$.  Since this map is $C^1$, $\abs{c-c'}\leq K_c \norm{u}_\h{1}$.\qed

If $\f_0$ is our perturbed $\f_c$, and $\norm{\f_0-\f_c}\leq \delta_{c}$, then we may apply Lemma \ref{lem:nearby-wave}, to find $\f_{c'}$ such that $\calN[\f_0] = \calN[\f_{c'}]$.  Since $\dist$ is a pseudo-metric on $\h{1}(\R)$,
\[
\dist(\f(t),\f_{c})\leq \dist(\f(t),\f_{c'}) + \dist(\f_{c},\f_{c'})
\]
$\dist(\f_{c},\f_{c'})$ is time independent and
\[
\dist(\f_{c},\f_{c'}) \leq \norm{\f_{c}-\f_{c'}}_\h{1}\leq C \abs{c-c'}\leq C \norm{\f_0-\f_c}_\h{1}
\]
which may be made arbitrarily small.  So it suffices to study the stability of $\f_{c'}$.

Examining how $\calN[\f]=\calN[\f_c]$ constrains the flow, first decompose $\f$ as
\begin{equation}
\f(x+x_0,t) = \f_c(x) + u(x,t)
\label{eq-decomposition}
\end{equation} 
the solitary wave and a perturbation, for some $x_0=x_0(t)$.  Expanding $\calN[\f]=\calN[\f_c]$ about $\f_c$, 
\[
\int_{-\infty}^{\infty} \bracket{I - \f_c^n \dx \paren{\f_c^n\dx \cdot} } \paren{\f_c-1} u dx=\inner{ \bracket{I - \f_c^n \dx \paren{\f_c^n\dx \cdot} }\paren{\f_c-1}}{u} = O(\norm{u}_\h{1}^{2})
\]
which we make formal in
\begin{lemma}
Given a solitary wave $\f_c$, there exists a constant $C$ such that for any $\norm{u}_\h{1}\leq \frac{1}{2}$,
\[
\mathcal{N}[\f_c+u]=\mathcal{N}[\f_c]
\]
\begin{equation}
\abs{\inner{ \bracket{I - \f_c^n \dx \paren{\f_c^n\dx \cdot}}\paren{\f_c-1}}{u}}\leq C \norm{u}_\h{1}^2
\label{eq:constraint1}
\end{equation}
\label{lem:invariantconstraint}
\end{lemma}
\begin{remark} Since the right-hand side of \eqref{eq:constraint1} is quadratic in $\norm{u}_\h{1}$, we view it as a \emph{near-orthogonality} constraint.\end{remark}

\subsubsection{The Sliding Metric and the Choice of $x_0(t)$}
\label{sec-sliding-metric}
As discussed in \cite{Bona75} with regard to the sliding metric, \eqref{eq:distance}, it is not true in general that the value of $y$ by which one function is translated to minimize the norm is be finite.  We will show, under some additional appropriate assumptions that this is the case, see \cite{Bona75, Bona94}.

\begin{lemma}
Given a solitary wave $\f_c$, assume that a solution to $\eqref{eq:magma}$ $\f$, $\f-1\in C^{1}([0,T):H^{1}(\mathbb{R}))$, satisfies
\begin{equation}
\dist(\f(t),\f_c) < \norm{\f_c-1}_\h{1}
\end{equation}
for $t<T$.  Then the infimum of the function $\rho$,
\begin{equation}
\rho(y) =  \norm{\f(\cdot)-\f_c(\cdot+y)}_\h{1}^2
\end{equation}
is achieved at a finite value of of $y\in \R$.
\label{lem:finitemin}
\end{lemma}
\proof$\rho(y)$ is obviously continuous, and because $\phi(\cdot+y)-1 \rightharpoonup 0$ in $H^1$ as $y\to \pm \infty$,
\[
\lim_{y\to \pm \infty} \rho(y) = \norm{\f-1}_\h{1}^2 + \norm{\f_c-1}_\h{1}^2 > \norm{\f_c-1}_\h{1}^2
\]
But by assumption,
\[
\inf_{y\in \R}\rho(y) = \inf_{y\in \R} \norm{\f(\cdot)-\f_c(\cdot+y)}_\h{1}^2 = d(\f,\f_c)^2 <  \norm{\f_c-1}_\h{1}^2
\]
So there must be some finite $y_0$ such that $\rho(y_0)<\rho(\pm\infty)$.  By the continuity of $\rho$, there then exists an $x_0(t)$ at each $t<T$ such that $\rho(-x_0(t))\leq \rho(y)$ for all $y\in\R$.\qed
\begin{remark}
There therefore exists a function $x_0(t)$ such that
\begin{equation}
\dist(\f,\f_c) = \norm{\f(\cdot,t)-\f_c(\cdot - x_0(t))}_\h{1}
\end{equation}
Since $\f_c$ is in fact smooth, so is $\rho(y)$, and hence $\rho'(-x_0(t))=0$,
\begin{equation}
\label{eq:constraint2}
\int_{-\infty}^{\infty} \bracket{I - \dx^2}\dx\f_c u dx=\inner{ \bracket{I - \dx^2}\dx\f_c }{u}=0
\end{equation}
where we used the decomposition \eqref{eq-decomposition}.  \eqref{eq:constraint2} is a second constraint on the perturbation to $\f_c$, which together with the near-orthogonality condition \eqref{eq:constraint1}, will be 
shown to yield local convexity of $\calE$ near $\f_c$.
\end{remark}

\subsection{Properties of the Linear Operator, $L_c$}
Here we summarize properties of $L_c$, and exhibit the non-positivity of $L_c$.
\begin{proposition}(Properties of the Linear Operator $L_c$)
\label{prop:Lc}
The linear second order operator defined by \eqref{eq:secondvariation} has the following features:
\begin{description}
  \item[\textbf{(i)}] $L_c$ is self adjoint, i.e. \begin{equation}L_c = L_c^{\dagger}\end{equation}
  \item[\textbf{(ii)}] $\dx\f_c$ is an eigenvector of $L_c$ with eigenvalue zero
  \begin{equation}L_c \dx \f_c = 0\end{equation}
  \item[\textbf{(iii)}] \begin{equation}L_c \partial_c \f_c = -\left[I - \f_c^n \dx \paren{\f_c^n\dx \cdot} \right]\paren{\f_c-1}= -c^{-1}\paren{\f_c^n -1}\end{equation}
  \item[\textbf{(iv)}] \begin{equation}\inner{L_c \partial_c\f_c}{\partial_c \f_c}=-\frac{d}{dc}\mathcal{N}[\f_c]\end{equation}
  \item[\textbf{(v)}] There exists $\lambda_0<0$ and $\psi_0 \in L^2(\R)$ such that
  \begin{equation}
  L_c \psi_0 = \lambda_0 \f_c^{2n} \psi_0
  \label{eq-generalized-evp}
  \end{equation}
  i.e. $\psi_0$ is the ground state of $L_c$ with eigenvalue $\lambda_0$ of the generalized eigenvalue problem \eqref{eq-generalized-evp}.
\end{description}
\end{proposition}
\proof
\textbf{(i-iv)} are trivial algebra and integration by parts.  For \textbf{(v)}, note that although $L_c$ is not in Sturm-Liouville form, if, given $h$, we let $\overline{h} = \f_c^n h$,  
\[
\f^{-n} L_c h = -c  \dx^2 \tilde{h} + \f_c^{-2n}\left[c - n\f^{n-1} -n c \f_c^{n-1}\dx\paren{\f_c^n \dx \f_c} \right] \tilde{h} = \overline{L}_c \overline{h}
\]
$\overline{L}_c$ is in standard Sturm-Liouville form, and it has a zero eigenvector, $\f_c^n \dx\f_c$.  Since this has one zero crossing, by oscillation theory (see \cite{Coddington84}, amongst others), we know there exists a ground state for $\overline{L}_c$, which we will denote by $\overline{\psi_0}$ with a negative eigenvalue, $\lambda_0$.  In turn, $L_c$ has a generalized eigenvector $\psi_0 = \f_c^{-n} \overline{\psi_0}$ with eigenvalue $\lambda_0$, so
\[
L_c \psi_0 = \lambda_0 \f_c^{2n} \psi_0
\]
We know this is the ground state of $L_c$ because if there existed some other $\psi$ with eigenvalue $\lambda < \lambda_0$, $\overline{\psi} = \f^n \psi$ would be an eigenvector of $\overline{L}_c$ with eigenvalue $\lambda$, which contradicts $\overline{\psi_0}$ being the ground state of $\overline{L}_c$.
\qed

We will now prove that with the constraints introduced in the previous section, \eqref{eq:secondvariation} admits the estimate
\[
\abs{\Delta \mathcal{E}} \geq \frac{1}{2}\inner{L_c u}{u}\geq C\norm{u}_\h{1}^2 + O\paren{\norm{u}_\h{1}^{3}}
\]
Defining the two vectors in $L^2$,
\begin{eqnarray}
\label{eq:xi1}
\xi_1 &=&\bracket{I - \f_c^n \dx \paren{\f_c^n\dx \cdot} }\paren{\f_c-1}\\
\label{eq:xi2}
\xi_2 &=&\bracket{I - \dx^2}\dx\f_c
\end{eqnarray}
\begin{proposition}
Let 
\begin{equation*}
\mathcal{A} = \left \{f \in \h{1}: \quad \norm{\f_c^n f}_{L^2}=1 \quad \mbox{and} \quad f \bot \xi_1 \right\}
\end{equation*}
Then
\begin{equation*}
\inf_{ f \in \mathcal{A} } \inner{L_c f}{f} =0
\end{equation*}
\end{proposition}

\proof
Following,\cite{Weinstein85, Weinstein86} , let $\alpha = \inf \inner{L_c f}{f}$, taken over $\mathcal{A}$.  Assume $\alpha<0$, and let us treat this as a constrained minimization problem.  From the theory of Lagrange multipliers, there exist $f_\star$, $\beta_\star$ such that
\begin{equation}
\paren{L_c - \alpha\f^{2n}}f_\star = \beta_\star \xi_1
\label{eq:variational-prob1}
\end{equation}
If $\beta_\star$ is zero, then $\alpha =\lambda_0$ which implies $f_\star$ is some multiple of the ground state $\psi_0$, as defined in Proposition \ref{prop:Lc}.  But since $\psi_0$ and $\xi_1$ are both even functions,
\[
\inner{f_\star}{\xi_1}  \propto \inner{\psi_0}{\xi_1} \neq 0
\]
and this contradicts the assumption that $f_\star$ is orthogonal to $\xi_1$.  Therefore, $\beta_\star \neq 0$.  If $\alpha = \lambda_0$, then, taking the inner product of both sides of \eqref{eq:variational-prob1} with the ground state,
\[
0 = \inner{f_\star} {\paren{L_c - \lambda_0\f^{2n}}\psi_0}= \inner{\paren{L_c - \alpha\f^{2n}}f_\star}{\psi_0} = \beta_\star \inner{\xi_1}{\psi_0} = \beta_\star \inner{ \frac{1}{c}\paren{\f_c^n-1}}{\psi_0}\neq 0
\]
So $\lambda_0 <\alpha<0$.

Let $g(\lambda)$ be defined as
\begin{equation*}
g(\lambda) = \inner{\paren{L_c - \lambda\f^{2n}}^{-1}\xi_1}{\xi_1}
\end{equation*}
on the interval $(\lambda_0,0]$.  Note
\begin{eqnarray*}
g'(\lambda) &=& \inner{\f_c^{2n}\paren{L_c - \lambda\f^{2n}}^{-1}\xi_1  }{\paren{L_c - \lambda\f^{2n}}^{-1}\xi_1}\\
&=&\norm{\f_c^n\paren{L_c - \lambda\f^{2n}}^{-1}\xi_1  }_{L^2(\R)}^2   >0
\end{eqnarray*}
so $g$ is increasing on this interval.  Additionally,
\begin{equation*}
g(0) = -\frac{d}{dc} \mathcal{N}[\f_c]<0
\end{equation*}
by Proposition \ref{prop:increasing-momentum}.  This implies that $g(\alpha)<0$.  But
\[
g(\alpha) = \frac{1}{\beta_\star}\inner{f_\star}{\xi_1} =0
\]
Therefore $\alpha=0$.
\qed

\begin{remark}
 It is here that we see the importance of the slope condition on $\calN$ with respect to $c$.  Also, we see from Figure \ref{fig-functional-manifold} (a) and (b), and the exact computation when $n=2$, that as $c\to n$, $\partial_c \calN[\f_c] \to 0$.  There is a bifurcation point at $c=n$, for there are not solitary waves for $c\leq n$, but under linearization about $\phi\equiv1$,there are plane waves with group velocity $\leq n$; this is cartooned in Figure \ref{fig-functional-manifold} (a).  The sign of the derivative of this functional with respect to $c$ was previously used in \cite{Barcilon89} to conclude linear \emph{instability} of one dimensional solitary waves in two spatial dimensions of \eqref{eq:magma}, when $n=3$ and $m=0$.
\end{remark}

\begin{proposition}
Let 
\begin{equation*}
\mathcal{B} = \left \{f \in \mathcal{A} : \quad   f \bot \xi_2 \right\}= \left \{f \in\h{1}: \quad  \norm{\f_c^nf}_{L_2}=1,\quad f\bot\xi_1,\quad f \bot \xi_2 \right\}
\end{equation*}
Then
\begin{equation*}
\inf_{ f \in \mathcal{B} } \inner{L_c f}{f} >0
\end{equation*}
\label{prop-quadratic-bound}
\end{proposition}
\proof
Let 
\[
\alpha = \inf_{ f \in \mathcal{B} } \inner{L_c f}{f}
\]
Since $\mathcal{B} \subset \mathcal{A}$, $\alpha \geq 0$ by the previous Proposition.  Assume that $\alpha = 0$ and this minimum is achieved at $f_\star$.  Again, by the theory of Lagrange multipliers,
\begin{equation*}
L_c f_\star = \beta_\star \xi_1 + \gamma_\star \xi_2
\end{equation*}
Taking the inner product of both $L_c f_\star$ with $\dx \f_c$, 
\begin{equation*}
0 = \inner{f_\star}{L_c \dx f_c} = \inner{L_c f_\star}{\dx f_c} = \beta_\star \inner{\xi_1}{\dx \f_c}+ \gamma_\star \inner{\xi_2}{\dx \f_c}
\end{equation*}
Since $\xi_1$ is an even function and $\dx \f_c$ is odd, $\inner{\xi_1}{\dx \f_c}=0$,  hence $\gamma_\star = 0$, implying
\begin{equation*}
f_\star = \kappa_\star \dx \f_c - \beta_\star \partial_c \f_c
\end{equation*}
Taking the inner product of $f_\star$ with $\xi_1$,
\begin{equation*}
0 = \inner{f_\star}{\xi_1}  = -\beta_\star \frac{d}{dc}\mathcal{N}[\f_c] \neq 0
\end{equation*}
Therefore $\alpha>0$.
\qed

\begin{corollary}
There exists a constant $C>0$ such that for all $f \in H^1(\R)$ orthogonal to both $\xi_1$ and $\xi_2$
\begin{equation*}
\inner{L_c f}{f} \geq C \norm{f}_{L^2}^2
\end{equation*}
which further implies
\begin{equation*}
\inner{L_c f}{f} \geq C' \norm{f}_\h{1}^2
\end{equation*}
\label{cor:lowerbound}
\end{corollary}
\proof
The first part is obvious by Proposition \ref{prop-quadratic-bound}.  To prove the second inequality, let us express $L_c$ using \eqref{eq:lin-op1} as
\[
L_c = - c \dx \paren{\f_c^{2n} \dx\cdot} + V(x)
\]
Then
\begin{eqnarray*}
\inner{L_c f}{f} &=&c \inner{\f_c^n \dx f}{\f_c^n \dx f} + \inner{Vf}{f} \\
&\geq& \norm{\dx f}_{L^2}^2 - \norm{V}_{L^\infty} \norm{f}_{L^2}^2\\
&\geq& \norm{\dx f}_{L^2}^2 - C^{-1} \norm{V}_{L^\infty}\inner{L_c f}{f}
\end{eqnarray*}
hence
\[
\inner{L_c f}{f} \geq \paren{1 + C^{-1}\norm{V}_{L^\infty}}^{-1} \norm{\dx f}_{L^2}^2
\]
and the inequality follows.
\qed

\begin{proposition}
Given a solitary wave $\f_c$ with $c>n>1$ for \eqref{eq:magma}, there exist positive constants $C_2$, $C_3$ such that for all $u\in \h{1}$ satisfying
\begin{eqnarray}
\label{eq:poly1}
\norm{u}_\h{1}&\leq& \frac{1}{2}\\
\label{eq:poly2}
\calN[\f_c +u] &=& \calN[\f_c]\\
\label{eq:poly3}
\inner{\xi_2}{u}&=&0
\end{eqnarray}
$\xi_2$ given by \eqref{eq:xi2}, we have
\begin{equation}
\inner{L_c u}{u} \geq C_2 \norm{u}_\h{1}^2 - C_3 \norm{u}_\h{1}^{3}
\label{eq:lowerbound}
\end{equation}
\label{prop:lowerbound}
\end{proposition}

\proof
Let $u= u_\bot + u_\|$ where
\begin{equation*}
u_\| = \inner{u}{\xi_1}\xi_1 
\end{equation*}
and
\begin{equation*}
u_\bot = u-u_\| = u-\inner{u}{\xi_1}\xi_1
\end{equation*}
Then
\begin{equation*}
\inner{L_c u}{u}  =  \inner{L_c\paren{u_\bot + u_\| }}{u_\bot +u_\|  } = \inner{L_c u_\bot}{u _\bot} + 2\inner{L_c u_\bot}{u_\| }+\inner{u_\| }{u_\| }
\end{equation*}
By Corollary \ref{cor:lowerbound} and Lemma \ref{lem:invariantconstraint},
\begin{eqnarray}
\nonumber\inner{L_c u_\bot}{u _\bot} &\geq& C\norm{u_\bot}_\h{1}^2 = C\bracket{\norm{u}_\h{1}^2-2\inner{u}{u_\| }-2\inner{\dx u}{\dx u_\| } +\norm{u_\| }_\h{1}^2}\\
\nonumber &=&C\left[\norm{u}_\h{1}^2-2\inner{u}{\xi_1}^2-2\inner{ \dx u}{\xi_1}\inner{u}{\xi_1} +\inner{u}{\xi_1}^2\norm{\xi_1}_\h{1}^2 \right]\\
 &\geq& C\norm{u}_{\h{1}}^2 - D\norm{u}_\h{1}^{3}
\label{eq:lowerbound1}
\end{eqnarray}

The other two terms follow more easily
\begin{eqnarray}
\nonumber\inner{L_c u_\|}{u_\|} &=& \inner{u}{\xi_1}^2\inner{L_c \xi_1}{\xi_1}\\ 
&\geq& -D\norm{u}_\h{1}^{4}
\label{eq:lowerbound2}
\end{eqnarray}

\begin{eqnarray}
\nonumber\inner{L_c u_\bot}{u_\|} &=& \inner{L_c u}{ u_\|} - \inner{L_c u_\|}{u_\|} = \inner{u}{\xi_1}\inner{u}{L_c  \xi_1}+\inner{u}{\xi_2}\inner{u}{L_c \xi_2}\\ 
&\geq& -D\norm{u}_\h{1}^{3}
\label{eq:lowerbound3}
\end{eqnarray}
\eqref{eq:lowerbound} then follows immediately from \eqref{eq:lowerbound1},\eqref{eq:lowerbound2}, and \eqref{eq:lowerbound3}.
\qed

\begin{remark}
Using the estimate \eqref{eq:lowerbound} together with \eqref{eq:energybound1},
\begin{eqnarray}
\abs{\Delta \calE} &\geq& \calE[\f_c +u]-\calE[\f_c]\geq \frac{1}{2}\inner{L_c u}{u}-D_3 \norm{u}_\h{1}^3\nonumber\\
 &\geq& \frac{1}{2}C_2 \norm{u}_\h{1}^2 - \paren{\frac{1}{2}C_2 +D_3}\norm{u}_\h{1}^3\equiv p_-\paren{\norm{u}_\h{1}}
\label{eq:lower-polynomial}
\end{eqnarray}
The polynomial $p_-$ will be important in the next section.
\end{remark}

\subsection{Proof of Orbital Stability and Global Existence in Time}
Unlike non-degenerate equations, such as KdV and BBM, we need to control $\norm{1/\f}_{L^\infty}$ in our existence proof, hence the additional condition appearing in Theorem \ref{thm:local-existence}.  It is a lack of \emph{a priori} bounds on this quantity that currently prevents a global existence proof for general $(n,m)$; this matter is discussed in \cite{Simpson07}.  However, we are able to prove, in tandem with nonlinear stability, global existence in time for data in a neighborhood of a solitary wave.

\subsubsection{Proof of Main Theorem}
\label{section:orbital-proof}
Let a particular solitary wave $\f_c$ be given for some $c>n >1$, and let $\f$ be a solution to \eqref{eq:magma}.  First we will consider the case where $\calN[\f]=\calN[\f_c]$; this will then be relaxed.

For $u$ satisfying \eqref{eq:poly1}, \eqref{eq:poly2}, and \eqref{eq:poly3}, the perturbation will satisfy the two inequalities
\[
p_-\paren{\norm{u}_\h{1}} \leq \abs{\Delta\calE}\leq p_+\paren{\norm{u}_\h{1}}
\]
$p_\pm$ defined in \eqref{eq:upper-polynomial} and \eqref{eq:lower-polynomial}.  Since $\Delta \calE$ is time independent, if the perturbation at time $t=0$ is sufficiently small, $\abs{\Delta\calE}$ may be made arbitrarily small.  Provided conditions \eqref{eq:poly1}, \eqref{eq:poly2}, and \eqref{eq:poly3} continue to hold, this will constrain $\norm{u}_\h{1}$ through $p_-$.

Let $\ep_\star$ be defined as
\begin{equation}
\ep_\star = \min \left\{\frac{1}{4}, \frac{1}{2}\norm{\f_c-1}_\h{1},\frac{C_2}{3\paren{C_3+2 D_3}} \right\}
\end{equation}
which depends only on $\f_c$.  The significance of the three quantities is:
\begin{itemize}
  \item $1/4$ will ensure $\f$ is bounded away from zero, as needed by our existence theory.
  \item $1/2 \norm{\f_c-1}_\h{1}$ will ensure the value at which the sliding metric is minimized is finite; see Lemma \ref{lem:finitemin}.
  \item $C_2/(3C_3 + 6 D_3)$ will ensure that the perturbation reamins to the left of the peak of the polynomial $p_-$.
\end{itemize}
Let $\ep\leq \ep_\star$ and let $\delta>0$ be sufficiently small such that
\begin{eqnarray}
\delta < \ep\\
p_+(\delta)< p_-(\ep)
\end{eqnarray}

Letting $\f(t=0)\equiv\f_0$, assume there is $x_0\in \R$ such that 
\[
\norm{\f_0-\f_c(\cdot+x_0)}_\h{1} =\norm{u_0}_\h{1}<\delta
\]  
$\f-1\in C^1\paren{[0,T_{\max});\h{1}}$.  With these choices of $\delta$, $\ep$, and $\ep_\star$, we will show that for $t\in [0,T_{\max})$, $\dist(\f(t),\f_c)<\ep$.

Let 
\[
\mathcal{I}=\left\{t: \dist(\f(t'),\f_c)<\eps\quad\mbox{for $t'< t$.}\right\}
\]
We will use $\mathcal{I}$ to prove the theorem by contradiction as follows:
\begin{itemize}
  \item Use continuous dependence upon the data of solutions of \eqref{eq:magma} to prove that $\mathcal{I}$ is not empty. 
  \item Seek the maximal time $T_0$ in $\mathcal{I}$. If it is not $T_{\max}$, we will show that there is some time interval beyond $T_0$ for which $\dist(\f(t),\f_c)<2\ep$.
  \item Prove that for any $t$ such that $\dist(\f(t),\f_c)<2\ep$, in fact $\dist(\f(t),\f_c)<\ep$, producing the contradiction.
\end{itemize}

First we prove that $\mathcal{I}$ is not empty.  This is an application of Theorem \ref{thm:local-existence}.
\[
\norm{\f_0-1}_\h{1} \leq \frac{3}{2}\norm{\f_c-1}_\h{1} \quad\mbox{and}\quad \norm{\frac{1}{\f_0}}_{L^\infty} \leq \frac{4}{3}
\]
Taking $R=3/2\norm{\f_c-1}_\h{1} $, $\alpha = 3/8$, we know from part (a) of Theorem \ref{thm:local-existence}, that $\norm{\f(t)-1}\leq 2 R$ and $\norm{1/\f(t)}_{L^\infty} \leq 1/\alpha$ up till at least $T_\exist (R, \alpha)>0$ and $T_{\max}\geq T_\exist$.

By part (c) of the same Theorem, we have a constant $K_\cont=K_\cont (R, \alpha)$, such that
\[
\norm{\f(\cdot,t)- \f_c(\cdot -ct +x_0)}_\h{1} \leq \norm{\f_0-\f_c(\cdot+x_0)}_\h{1} e^{K_\cont t}< \delta  e^{K_\cont t}
\]
Taking $t$ sufficiently small, there is some time interval over which $\f$ is within $\ep$ of $\f_c$.  

Let
\begin{equation}
T_0 = \sup \mathcal{I}
\end{equation}
Suppose $T_0 < T_{\max}$.  For any $t< T_0$, 
\[
\dist(\f(t),\f_c)<\eps\leq  \frac{1}{2}\norm{\f_c-1}_\h{1}
\]
Lemma \ref{lem:finitemin} asserts there exists $x_0(t)\in \R$ such that
\[
\dist(\f(t), \f_c)= \norm{\f(\cdot,t)-\f_c(\cdot +x_0(t))}_\h{1}< \ep
\]
Furthermore, for all $t< T_0$, we have again
\[
\norm{\f(t)-1}_\h{1} \leq \frac{3}{2}\norm{\f_c-1}_\h{1}\quad \mbox{and}\quad \norm{\frac{1}{\f(t)}}_\h{1} \leq \frac{4}{3}
\]
We use these bounds to control how far $\f$ can deviate from $\f_c$ beyond $T_0$.  Taking $R=3/2 \norm{\f_c-1}_\h{1}$ and $\alpha= 3/8$, there exists $T_\exist (R, \alpha)>0$ such that $\f(t)$, for $t< T_0$, may be continued in time by the amount $T_\exist$ and will satisfy $\norm{\f(t)-1}\leq 2R $, and $\norm{1/\f(t)}_{L^\infty} \leq 1/\alpha$.  Because the solution is unique, these bounds apply to $\f$ for $t< T_0+ T_\exist$.

With this control on norms, we let  $t_1< T_0$ be a new starting point, and apply part (c) of Theorem \ref{thm:local-existence}
\[
\norm{\f(\cdot, t)-\f_c(\cdot -c t+ x_0(t_1))}_\h{1}\leq \norm{\f(\cdot,t_1) - \f_c(\cdot+x_0(t_1)}_\h{1}e^{K_\cont (t-t_1)}< \eps e^{K_\cont (t-t_1)}
\]
Making $t_1$ sufficiently close to $T_0$, we can find $T_2 > T_0$ for which 
\[
\norm{\f(\cdot, t)-\f_c(\cdot -c t+ x_0(t_1))}_\h{1}=\norm{u(t)}_\h{1}< 2 \ep
\]
for $t< T_2$.  We claim this implies the stricter estimate $\norm{\f(\cdot, t)-\f_c(\cdot -c t+ x_0(t_1))}_\h{1}<\ep$ for $T_0<t<T_2$.

$2\ep\leq 2\ep_\star \leq \norm{\f_c-1}_\h{1}$; applying Lemma \ref{lem:finitemin} to find $x_0(t)$ again, and then decompose our solution via \eqref{eq-decomposition}.  As noted in the remark following Lemma \ref{lem:finitemin}, this perturbation satisfies $\inner{\xi_2}{u}=0$.  $\calN[\f] = \calN[\f_c]$ and 
\[
\norm{u}_\h{1}<2\ep\leq 2\ep_\star \frac{1}{2}
\] 
Therefore $u$ satisfies the neccessary conditions to apply Proposition \ref{prop:lowerbound}, 
\[
p_-\paren{\norm{u}_\h{1}}\leq \abs{\Delta \calE}\leq p_+(\delta)< p_-(\ep)
\]
Because $\norm{u}_\h{1}< 2\eps\leq 2\ep_\star \leq 2C_2/(3 C_3 + 6 D_3)$, it sits remains to the left of the peak of $p_-$;
\[
\norm{u(t)}_\h{1}=\dist(\f(t),\f_c) < \ep
\]
This holds for $T_0< t<T_2$, $T_0< T$ so we have a contradiction and conclude $\dist(\f(t),\f_c)<\ep$ for all $t< T$.

Now we relax $\calN[\f_0] = \calN[\f_c]$.  First apply Lemma \ref{lem:nearby-wave} to $\f_c$, to find $\delta_c$ that will define the $\h{1}$ neighborhood about the origin where the initial perturbation must reside.  Let $K$ be the constant such that
\[
\norm{\f_c - \f_{c'}}_\h{1} \leq \tilde K \abs{c-c'}\leq \tilde K K_c \norm{u}_\h{1}=K \norm{u}_\h{1},
\]
where $K_c$ is from Lemma \ref{lem:nearby-wave}.  We will seek a new wave $\f_{c'}$ with which to apply the preceding argument; however, $\ep_\star$ and the polynomials $p_\pm$ will be determined by $\f_{c'}$.  But $\f_{c'}$ is determined by $\f_0$ and we have not yet found \emph{all} the bounds $\f_0$ must satisfy.  Uniform control is needed.  Using $\mathcal{G}$, the implicit function associated with Lemma \ref{lem:nearby-wave}, let 
\begin{eqnarray*}
c_{\min} &=& \min_{\norm{u}_\h{1}<\delta_c} \mathcal{G}[u]\\
c_{\max} &=& \max_{\norm{u}_\h{1}<\delta_c} \mathcal{G}[u]
\end{eqnarray*}
and let 
\[
\ep_\star' = \min_{c\in [c_{\min},c_{\max}]} \ep_\star(\f_c)
\]
Similarly, the coefficients of $p_\pm$ may be chosen such that
\[
p'_-\paren{\norm{u}_\h{1}} \leq \abs{\calE[\f_c+u]-\calE[\f_c]}=\abs{\Delta \calE} \leq p'_+\paren{\norm{u}_\h{1}}
\]
for all $c \in [c_{\min},c_{\max}]$, $\norm{u}_\h{1} < \min\{\delta_c,\frac{1}{2}\}$ (the $\frac{1}{2}$ is required by Lemma \ref{lem:energybound}).

Given $\f_c$, let $\ep_\star'$ be as above.  Let $\ep_\star = \ep_\star' (1+K)$.  Take $\ep\leq \ep_\star$, and then set $\ep' = \ep/(1+K)$.  Choose $\delta$ such that 
\begin{eqnarray}
\delta &<& \delta_c\\
(1+K)\delta &< &\ep'\\
p'_+((1+K)\delta) &<&p'_-(\ep')
\end{eqnarray}
and let $\delta' = \delta(1+K)$.

Assume
\[
\norm{\f_0-\f_c(\cdot+x_0)}_\h{1}<\delta
\]
Let $\f_{c'}$ be the neighboring solitary wave for which $\calN[\f_0]=\calN[\f_{c'}]$.  Note $\f_0$ is also close to $\f_{c'}$,
\[
\norm{\f_0(\cdot)-\f_{c'}(\cdot+x_0)}_\h{1}\leq \norm{\f_0(\cdot)-\f_c(\cdot+x_0)}_\h{1}+ \norm{\f_{c'}-\f_c}_\h{1}< (1+K)\delta = \delta'
\]
Because $\delta' < \ep'$ and $p'_+(\delta') < p'_-(\ep')$, $\ep'\leq \ep_\star'$, we may apply the previous argument, to conclude
\[
\dist(\f(t),\f_{c'})< \ep'
\]
for $t \in [0, T_{\max})$.
Finally,
\[
\dist(\f(t),\f_c)\leq \dist(\f(t),\f_{c'}) + \dist(\f_{c'},\f_c)< \eps' +K \delta\leq\eps' +K \delta' < (1+K) \ep'=\ep  
\]
\qed
\subsubsection{Proof of Corollary \ref{cor:globalexist}}
Now we will prove global existence in time for a data in a neighborhood of a solitary wave.  Given a solitary wave $\f_c$, $\ep\leq \ep_\star$, and any function $\f$ such that
\[
\norm{\f_0(\cdot)-\f_c(\cdot-x_0)}_\h{1} < \delta
\]
Let $T_{\max}$ be the maximal time of existence of $\f$, the solution emanating from $\f_0$.  Suppose $T_{\max}<\infty$.

By Theorem \ref{thm:orbital}, for all $t<T_{\max}$,
\[
\dist(\f(t),\f_c) <\ep
\]
which implies
\begin{eqnarray*}
\f(t) > 1-\ep\geq 1/2\\
\norm{\f(t)-1}_\h{1} < \ep + \norm{\f_c-1}_\h{1} < 2 \norm{\f_c-1}_\h{1}
\end{eqnarray*}
and hence
\[
\norm{\f-1}_\h{1} +\norm{\frac{1}{\f}}_{L^\infty} < \frac{1}{2 \norm{\f_c-1}_\h{1}}+2
\]
for all $t< T_{\max}$.  But $\f$ is $C^1$ in time, and since $T_{\max}<\infty$,
\[
\lim_{t\to T_{\max}} \norm{\f-1}_\h{1} +\norm{\frac{1}{\f}}_{L^\infty}=\infty
\]
a contradiction.  Therefore $T_{\max} = \infty$, and furthermore, by Theorem \ref{thm:orbital} again,
\[
\dist(\f(t),\f_c) <\ep
\]
for all time.\qed

\section{Relation to Compacton Equations}
\label{sec:compacton}
Compactons, robust compactly supported solitary waves, were first identified in \cite{Rosenau93} in a generalization of KdV, $K(m,n)$,
\[
\dt u + \dx \paren{u^m} + \frac{1}{n} \dx^3\paren{u^n}=0
\]
$K(m,n)$ has been further generalized to $C_1(m, a+b)$, 
\begin{equation}
\dt u + \dx \paren{u^m}+\frac{1}{b} \dx \bracket{u^a \dx^2 \paren{u^b}}=0
\label{eq:c1}
\end{equation}
discussed in \cite{Rosenau06a}, which are known admit traveling wave solutions of speed $c\geq 0$, compactly supported in space.  In turn, this has been generalized to the multidimensional case, $C_N(m, a+b)$, in \cite{Rosenau06a, Rosenau07}.

\begin{figure}
\begin{center}
\includegraphics[width=4in]{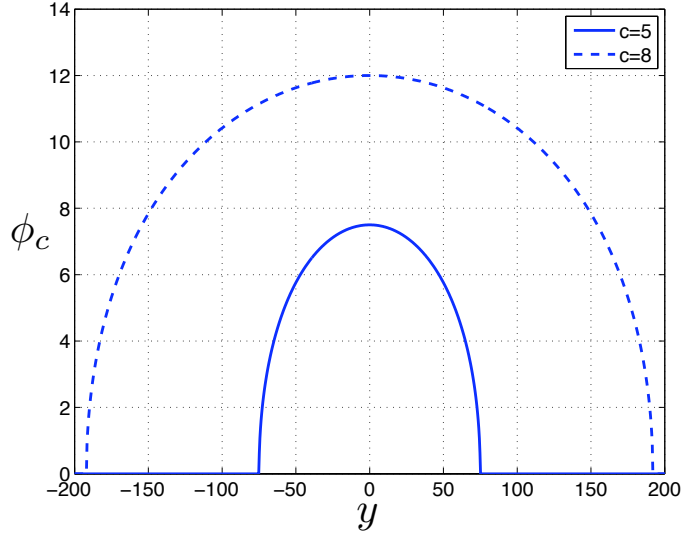}
\caption{Compacton solutions to \eqref{eq:magma} with $n=2$, $m=-2$ for $c=5$ and $c=8$.  Faster compactons are both broader and taller.}
\label{fig:compacton}
\end{center}
\end{figure}

Like \eqref{eq:magma}, \eqref{eq:c1} is a nonlinearly dispersive wave equation.  Indeed, \eqref{eq:magma} may be written as
\begin{equation}
\label{eq:magma-compacton-version}
\dt \phi + \dx\paren{\phi^n} - \frac{1}{1-m}\dx \bracket{\phi^n \dx\dt \paren{\phi^{1-m}}}=0
\end{equation}
which, upon setting $a=n$ and $b=1-m$, resembles \eqref{eq:c1} up to one $\partial_x$ becoming $-\partial_t$.  It was the recognition of this similarity between these two equations that led the authors to discover that just as \eqref{eq:c1} is Hamiltonian for $ b-a=1$, so too is \eqref{eq:magma} when $ 1-m - n = 1$, $n+m=0$.  \eqref{eq:magma} may be interpreted as a generalization of the BBM equation, just as \eqref{eq:c1} is a generalization of KdV.  In addition, \eqref{eq:magma} also possesses compacton solutions, pictured in Figure \ref{fig:compacton}.

Equation \eqref{eq:c1} has a second set of Hamiltonian cases when $ b-a = \frac{1}{2}$, which is related to the property that the density of the generalized momentum of \eqref{eq:c1}
\[
\int u^{1 + b-a}
\]
is a mapping of a \eqref{eq:c1} into another such equation.  The generalized momentum of \eqref{eq:magma}, \eqref{eq:momentum}, not being a monomial, lacks this property.  Note also that \eqref{eq:magma} could be further generalized by letting the the nonlinearity in the $(\phi^n)_x$ vary independently of the dispersive term; however, for physical reasons related to its derivation, we leave it as is.

We now construct compactons.  Starting with \eqref{eq:magma-compacton-version}, we integrate it up, assuming there exists $y_{\max} >0$, such that for $\abs{y}> y_{\max}$, $Q_c=0$.  Then
\begin{equation}
-c Q_c + Q_c ^n + \frac{c}{1-m}Q_c^n\partial_y^2 \paren{Q_c ^{1-m}}=0 \quad\mbox{for $\abs{y}\leq y_{\max}$}
\label{eq:compacton1}
\end{equation}
with boundary conditions $Q_c = 0$ at $\abs{y} = y_{\max}$.

Introducing the scaling
\begin{equation}
Q_c(y) = c^{\frac{1}{n-1}}U(\sqrt{1-m} c^{-\frac{n-m}{2(n-1)}} y)^{\frac{1}{1-m}} =  c^{\frac{1}{n-1}}  U(\xi)^{\frac{1}{1-m}}
\label{eq:compacton-scaling}
\end{equation}
into \eqref{eq:compacton1}, $U$ solves
\begin{equation}
-U^{-\frac{n-1}{1-m}} +1 + \partial_\xi^2 U = 0
\label{eq:U-BVP}
\end{equation}
\begin{remark}
Assuming $n>1$, the scaling, \eqref{eq:compacton-scaling}, highlights three distinct regimes
\begin{itemize}
  \item If $n>m$, faster waves are taller and narrower.
  \item If $n<m$, faster waves are taller and broader.
  \item If $n=m$, faster waves are taller, but all waves have the same width.
\end{itemize}
\label{rem:compacton-scaling}
\end{remark}

Integrating \eqref{eq:U-BVP} again,
\[
\frac{1}{2}\paren{\partial_\xi U}^2 + \mathrm{V}(U; \beta)= 0
\]
where the potential, $\mathrm{V}$,
\[
\mathrm{V}(U;\beta )= \left\{ 
\begin{array}{cc}
U - \frac{1}{1-\beta}U^{1-\beta},&\mbox{ if } \beta \neq 1\\
U-\log U, &\mbox{ if } \beta = 1
\end{array}\right.
\]
and $\beta = (n-1)/(1-m)$.  

\begin{figure}
\begin{center}
\includegraphics[width=3in]{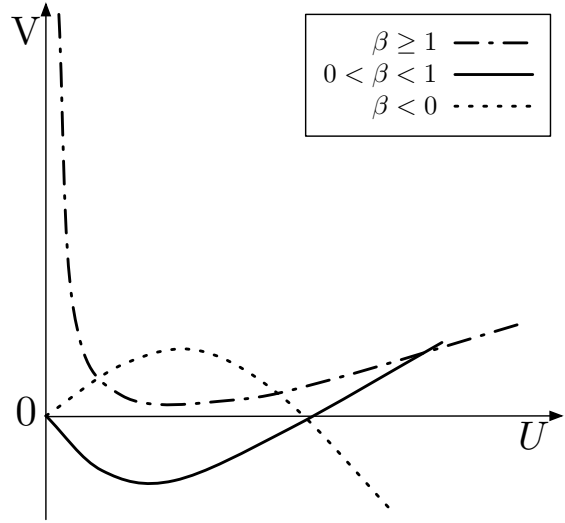}
\caption{The different regimes of $\mathrm{P}$ for different $\beta$.  We exclude $\beta=0$ because $\mathrm{P}(\beta=0)\equiv 0$.}
\label{fig:beta-plot}
\end{center}
\end{figure}

Representative cases for $\beta \geq 1$, $0<\beta<1$ and $\beta<0$ are illustrated Figure \ref{fig:beta-plot}.  Compactons are homoclinic orbits in the $(U,U')$ phase plane connecting the equilibrium point $(0,0)$ to itself in finite time.  This corresponds to $\mathrm{V}$ having a potential well between $U=0$ and some $U_{\max}>0$.  Such a well exists only for $0<\beta<1$; see Figure \ref{fig:beta-plot}.

In terms of $n$ and $m$, for $n>1$,  $0<\beta<1$ corresponds to $n+m <2$, which obviously includes the Hamiltonian case of \eqref{eq:hamiltonian-form}, \eqref{eq:hamiltonian}, \eqref{eq:skewsym-op}; $n+m=0$.  With regard to Remark \ref{rem:compacton-scaling}, because we have assumed $n>1$, we are in regime where faster waves are both taller \emph{and} broader.  

We note that several examples of compactly supported solutions of \eqref{eq:magma} were previously examined in \cite{takahashi1988esm,takahashi1990pma}.  These included the exponents $(n,m) = (3,0), (4,0)$.  The authors argued against the realization in nature of such solutions due to a stress singularity.

Some solutions of \eqref{eq:U-BVP}, given implicitly by
\begin{eqnarray}
-\frac{24 U+10 U^{4/5}+50 U^{3/5}-125 U^{2/5}}{6 \sqrt{10
   U^{4/5}-8 U}}= \abs{\xi}\quad\mbox{for $\beta = \frac{1}{5}$}\\
\frac{-2 U-3 U^{2/3}+9 \sqrt[3]{U}}{\sqrt{3 U^{2/3}-2 U}} = \abs{\xi} \quad\mbox{for $\beta=\frac{1}{3}$}\\
-\frac{\sqrt[4]{U} \left(2 \sqrt{\sqrt{U}-2} \log
   \left(\sqrt{\sqrt{U}-2}+\sqrt[4]{U}\right)+U^{3/4}-2
   \sqrt[4]{U}\right)}{\sqrt{\sqrt{U}-U/2}}= \abs{\xi}\quad\mbox{for $\beta=\frac{1}{2}$}
\end{eqnarray}
and 
\[U\in \paren{0, \paren{\frac{1}{1-\beta}}^{\frac{1}{\beta}}}
\]

The compactons are a type of \emph{weak} solution to \eqref{eq:magma}, but the exact notion is still imprecise.  Near the left edge of the compacton, $U\sim \abs{\xi+\xi_{\max}}^{2/(1+\beta)}H(\xi+\xi_{\max})$, $H(x)$ the heavyside function.  In the Hamiltonian case $Q_c \sim \abs{y+y_{\max}}^{1/n}H(y+y_{\max})$.  For $n\geq 2$, this will not have a square integrable derivative.  They do satisfy the following definition, previously given in \cite{Simpson07} for solutions of 
\eqref{eq:magma} that go to zero.

\begin{definition}
$\f(x,t)$ is a solution of \eqref{eq:magma} if
\begin{eqnarray}
\int_0^\infty \int_{-\infty}^{\infty} &\Big[&-\dt\psi(x,t) \f(x,t) -\dx\psi(x,t) \f(x,t)^n\nonumber\\
&& + \frac{1}{1-m}\dx \psi(x,t)\f(x,t)^n \dx\dt\paren{ \f(x,t)^{1-m} }\Big]dx dt =0
\label{eq:weaksolution}
\end{eqnarray}
for all $\psi(x,t) \in C^{\infty}_0(\R \times \R^+)$, and such that both
\begin{equation*}
\f(x,t) \quad\mbox{and}\quad \f(x,t)^n \dx\dt\paren{ \f(x,t)^{1-m} }
\end{equation*}
are in $L^1_{loc}(\R)$ in the $x$ coordinate.
\label{def:weak-solution}
\end{definition}

\begin{proposition}
$Q_c$ is a solution of \eqref{eq:magma} in the sense of Definition \ref{def:weak-solution}.
\end{proposition}
\proof
For $\abs{x-ct} > y_{\max}$, $Q_c \equiv 0$, and satisfies \eqref{eq:compacton1} pointwise.  For $\abs{x- ct} < y_{\max}$, $Q_c$ is smooth and solves \eqref{eq:compacton1} in the classical sense.  Now consider $y=x-ct$ near $-y_{\max}$.  In this neighborhood, $Q_c^n\partial_y^2 \paren{Q_c^{1-m}} \sim \abs{y+y_{\max}}^{2/(n-m)}H(y+y_{\max})$.  So for $n-m>0$, this is a continuous function and \eqref{eq:compacton1} also holds pointwise.  Therefore,
\begin{eqnarray*}
\int_0^\infty \int_{-\infty}^{\infty} &\Big [&-\dt\psi(x,t) Q_c(x-ct) -\dx\psi(x,t) Q_c(x-ct)^n \\
&-& \frac{c}{1-m}\dx \psi(x,t)Q_c(x-ct)^n \dx^2\paren{ Q_c(x-ct)^{1-m} }\Big ]dx dt \\
&=&\int_0^\infty \int_{-\infty}^{\infty} \paren{-\dt\psi(x,t)-c\dx\psi(x,t)} Q_c(x-ct) dx dt
\end{eqnarray*}
But this is the weak form of the transport equation $\dt u + c \dx u=0$, which $Q_c$ solves.  So the integral is zero for all test functions and $Q_c$ is a weak solution in this sense.
\qed

Finally, if instead of having the Hamiltonian as in \eqref{eq:hamiltonian}, we set
\[
\mathcal{H} = \int \paren{ -\frac{1}{n+1}\f^{n+1} }dx
\]
and replace the generalized momentum $\calN$ with
\[
\calN = \int \paren{\frac{1}{2}\f^{2n}\paren{\dx \f}^2 +\frac{1}{2}\f^2 }dx
\]
then these formally conserved quantities are finite for the compactons.  The compactons are then critical points of the energy functional $\mathcal{E} = \mathcal{H} + c \mathcal{N}$.  This suggests the possibility of an analogous stability argument as for the solitary waves.  One can take the second variation and formulate the spectral problem
\[
L_c u =-c \dx \paren{{Q_c}^{2n} \dx u} - \bracket{(2n-1) n c Q_c ^{2n-2}(\dx Q_c)^2+2 n c{Q_c}^{2n-1} \dx ^2 Q_c +n {Q_c}^{n-1} -c}u
\]
just as in Section \ref{sec:variational}.  However, a well-posedness theorem for solutions of \eqref{eq:magma} that vanish outside a compact set must be formulated before this is pursued.

\section{Remarks and Open Questions}
We have presented a new class of Hamiltonian PDEs with orbitally stable solitary waves.  A consequence of our stability analysis is the extension of well-posedness results for this system in \cite{Simpson07}  to a neighborhood of any solitary wave.   As noted, except for the case $n=2$, this result is currently only valid up to the acceptance of numerical computation and estimation of the slope of the invariant $\calN[\f_c]$.  We also observed that our equations have {\it compacton} solutions. These are solitary traveling waves, whose spatial support is compact. We show that these compactons solve the evolution equation in  a weak sense.

Compactons warrant further examination.  Formally, compactons are critical points of the functional $\mathcal{E}[\phi]$, defined in section \ref{sec:compacton}. A well-posedness theory in a function space, with respect to which the mapping $\phi\mapsto\mathcal{E}[\phi]$ is continuous,  and a spectral analysis of the second variation about a compacton analogous to that for the solitary waves of section \ref{sec-stability-proof}, which would imply stability of compactons. This is an interesting open problem.

\section*{Acknowledgements}
We thank Marc Spiegelman for his helpful comments and support, in addition to his contributions through the results appearing in \cite{Simpson07}.

\thanks{This work was funded in part by the US National Science Foundation (NSF) Collaboration in Mathematical Geosciences (CMG), Division of Mathematical Sciences (DMS), Grant DMS-0530853, the NSF Integrative Graduate Education and Research Traineeship (IGERT) Grant DGE-0221041, NSF Grants DMS-0412305 and DMS-0707850 and the Israeli Science Foundation Contract 801/07}

\appendix
\section{Generalized Momentum}
\label{sec:momentum}
The Lagrangian density, $\mathcal{L}$, for the Hamiltonian case of \eqref{eq:magma} is given by $\psi_x = \phi-1$,
\begin{equation}
\mathcal{L}(\psi, \psi_t) = \frac{1}{2}\psi_t\psi_x+ \frac{1}{n+1}\paren{(\psi_x+1)^{n+1}-1}-\psi_x + \frac{1}{2} \paren{\psi_x+1}^{2n} \psi_{xx} \psi_{tx}
\end{equation}
\eqref{eq:magma} is then Euler-Lagrange equation corresponding to $\delta \int \int \mathcal{L} dx dt$. 
\begin{equation}
\frac{\delta \mathcal{L}}{\delta \psi_t} = \frac{1}{2}\psi_x - \paren{ \frac{1}{2} \paren{\psi_x+1}^{2n} \psi_{xx}}_x = \pi
\end{equation}
and 
\[
\inner{\pi}{\psi_x} = \mathcal{N}[\f]
\]
the generalized momentum, \eqref{eq:momentum}, of \eqref{eq:magma} in the case $n+m=0$.
\bibliographystyle{plain}
\bibliography{hamiltonian_article}

\end{document}